\documentclass[aps,prl,twocolumn,amsmath,amssymb]{revtex4-1}
\usepackage{tabularx}
\usepackage{bm}
\usepackage{euscript}
\usepackage{graphicx}
\usepackage{color}
\usepackage{amsfonts}
\usepackage{exscale}
\usepackage{amsbsy}
\usepackage{subfigure}
\usepackage{textcomp}

\begin{document}

\title{Conductance and noise signatures of Majorana backscattering}

\author{Suk Bum Chung$^1$, Xiao-Liang Qi$^{1,2}$, Joseph Maciejko$^1$, and Shou-Cheng Zhang$^1$}
\affiliation{$^1$Department of Physics, Stanford University, Stanford, CA 94305\\
               $^2$ Microsoft Research, Station Q, Elings Hall, University of California, Santa Barbara, CA 93106}
\date{\today}

\begin{abstract}
We propose a conductance measurement to detect the backscattering of chiral Majorana edge states. Because normal and Andreev processes have equal probability for backscattering of a single chiral Majorana edge state, there is qualitative difference from backscattering of a chiral Dirac edge state, giving rise to half-integer Hall conductivity and decoupling of fluctuation in incoming and outgoing modes. The latter can be detected through thermal noise measurement. These experimental signatures of Majorana fermions are robust at finite temperature and do not require the size of the backscattering region to be mesoscopic.
\end{abstract}

\maketitle

From particle physics to condensed matter physics, Majorana fermions currently arouse great interest~\cite{WILCZEK2009}. In particle physics, where the concept first originated~\cite{Majorana1937}, the experimental signature of Majorana fermions, though not yet detected, has been known for a long time, i.e. the neutrinoless double $\beta$ decay in the case of neutrinos.

In condensed matter physics, two-dimensional (2D) systems where Majorana fermions can arise have been attracting a great deal of attention recently, partly due to potential applications to topological quantum computation~\cite{MOORE1991, READ2000, Ivanov2001, KITAEV2006, NAYAK2008, Fu2008}. One class of systems where Majorana fermions can appear is the 2D chiral superconductor which has a full pairing gap in the bulk, and $\mathcal{N}$ gapless chiral 1D Majorana fermions~\cite{Volovik1988, READ2000} at the edge. This system can be considered as the superconducting analogue of the quantum Hall (QH) state with $\mathcal{N}$ gapless chiral edge states, and is called a topological superconductor (TSC)~\cite{Qi2009}.

The challenge at hand is to find a way to detect the Majorana nature of the 1D edge state in this class of systems. So far, there has been no experiment which explicitly shows the Majorana nature of gapless states at the boundary of a TSC, though methods of detection have been proposed recently~\cite{Chung2009,fu2009d}. The first issue is to find a physical system which unequivocally belongs to this class. Despite theoretical prediction that the superconducting phase of Sr$_2$RuO$_4$ is the spinful version of the $\mathcal{N}=1$ chiral TSC due to its $p_x + ip_y$ pairing~\cite{Rice1995}, attempts to detect the gapless edge states have note been successful~\cite{Kam-edge}. 
Therefore, our discussion will be based on a recent proposal to induce topological superconductivity through the proximity effect on a magnetically doped Bi$_2$Se$_3$ film~\cite{Qi2010}. The second issue is to devise an experiment that can work at reasonable temperatures, as interference effects get washed out above the temperature scale set by the size of the TSC region~\cite{FU2009,akhmerov2009}. Our approach makes use of a generalization~\cite{ANANTRAM1996} of the Landauer-B\"{u}ttiker formalism~\cite{BUTTIKER1992} to superconducting systems, which was recently used 
to study the detection of a zero-dimensional Majorana bound state~\cite{NILSSON2008}.

In this Letter, we study the backscattering of the edge state of a $\mathcal{N}=1$ quantum anomalous Hall (QAH) state off a TSC island. We find a strikingly different behavior in both conductance and noise depending on the topological invariant $\mathcal{N}$ of the TSC, which is equal to the number of chiral Majorana edge states. In particular, for strong backscattering by the $\mathcal{N}=1$ TSC, the incoming and outgoing channels in the leads decouple because the probabilities for normal and Andreev scattering become equal. Indeed, whereas Andreev processes do not play any role in the case of the normal, topologically trivial superconductor (NSC) or the $\mathcal{N}=2$ TSC, backscattering due to the $\mathcal{N}=1$ TSC imposes a special condition between the probabilities for normal and Andreev scattering. The reason why Andreev scattering can occur for the $\mathcal{N}=1$ TSC is that in this case, the single chiral Dirac edge state of the QAH state splits into {\it two} chiral Majorana edge states. We emphasize that this is a pure transport experiment that can detect the Majorana nature. 

\begin{figure}[t]
\centerline{\includegraphics[width=.3\textwidth]{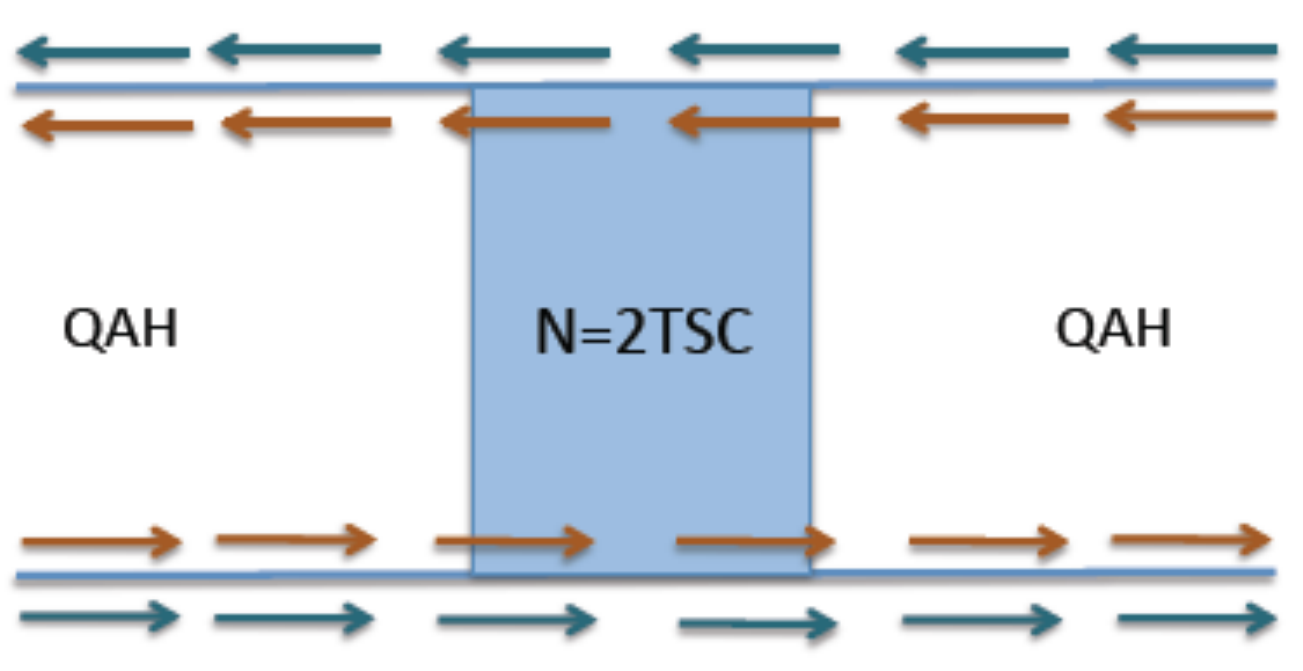}}
\medskip \medskip \medskip
\centerline{\includegraphics[width=.3\textwidth]{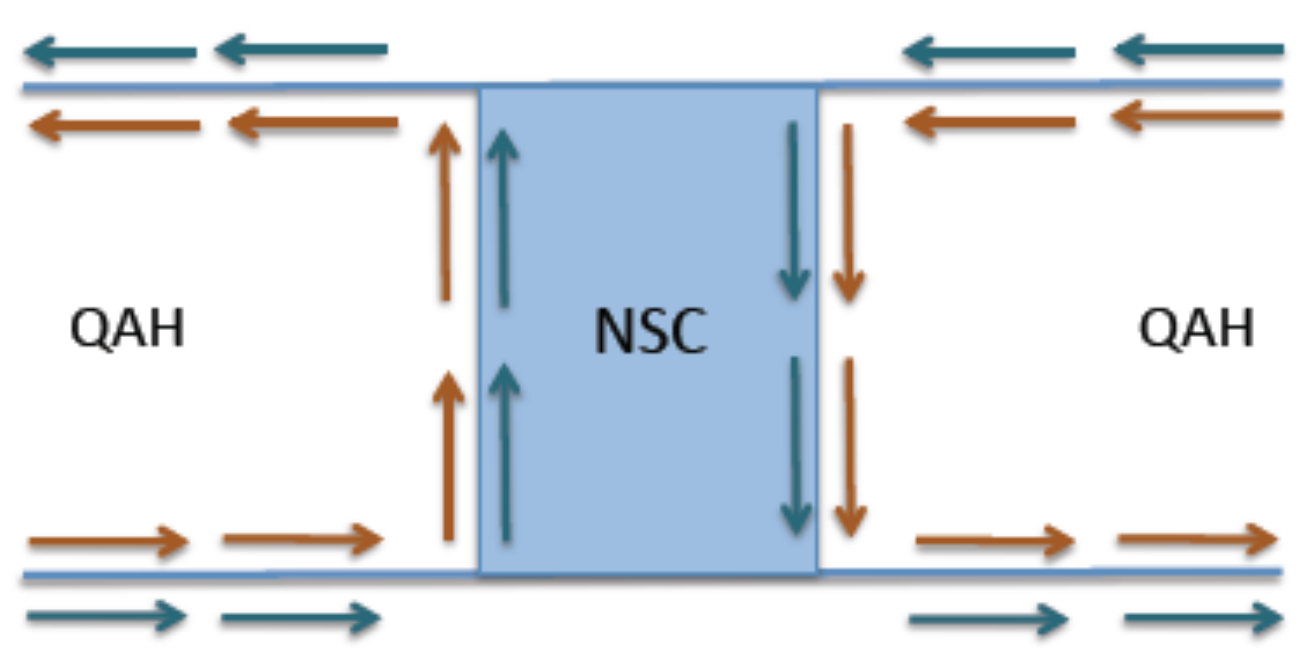}}
\medskip \medskip \medskip
\centerline{\includegraphics[width=.3\textwidth]{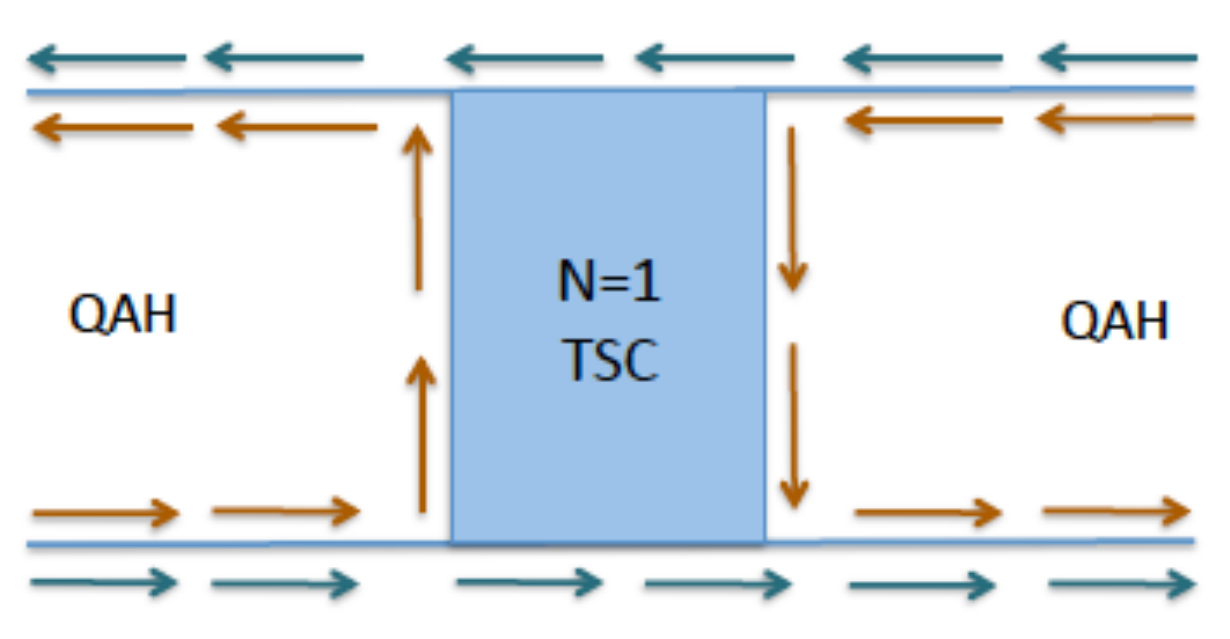}}
\medskip
\caption{Comparison between different SC. There is no backscattering for $\mathcal{N}=2$ TSC, Dirac fermion backscattering for NSC, and Majorana fermion backscattering for $\mathcal{N}=1$ TSC. Red and blue arrows represent $e\pm h$ chiral Majorana edge states, respectively.}
\label{FIG:QAH-TSC backscatter}
\end{figure}

We first discuss how to obtain a $\mathcal{N}=1$ TSC. When Cr or Fe magnetic dopants are introduced into Bi$_2$Se$_3$ or Bi$_2$Te$_3$ thin films, the spin exchange interaction leads to the effective Hamiltonian~\cite{Yu2010}
\begin{equation}\nonumber
h_\mathrm{QAH} = \left(\begin{array}{cc}m+Bp^2 & A(p_x -ip_y)\\ A(p_x + ip_y) & -m-Bp^2\end{array}\right),
\end{equation}
where the basis is $(c_{{\bf p}\uparrow}, c_{{\bf p}\downarrow})^T$ with $c_{\mathbf{p}\sigma}$ annihilating an electron of momentum $\mathbf{p}$ and spin $\sigma=\uparrow,\downarrow$ and 
the QAH effect obtained when $m<0$. The proximity effect gives us the Bogoliubov-de Gennes (BdG) Hamiltonian
\begin{equation}\nonumber
h_\mathrm{BdG} = \left(\begin{array}{cc} h_\mathrm{QAH}({\bf p})-\mu & i\Delta\sigma^y\\ -i\Delta^*\sigma^y & -h^*_\mathrm{QAH}(-{\bf p})+\mu\end{array}\right),
\end{equation}
the basis for which is $(c_{{\bf p}\uparrow}, c_{{\bf p}\downarrow},c^\dagger_{-{\bf p}\uparrow}, c^\dagger_{-{\bf p}\downarrow})^T$. A particularly simple case is $\mu = 0$, for which we can rewrite the BdG Hamiltonian as
\begin{equation}
h_\mathrm{BdG} = \left(\begin{array}{cc} h_+({\bf p}) & 0 \\ 0 & -h^*_-(-{\bf p})\end{array}\right),
\label{EQ:zeroChem}
\end{equation}
where
\begin{equation}
h_\pm ({\bf p}) = \left(\begin{array}{cc} m\pm|\Delta|+Bp^2 & A(p_x - i p_y)\\ A(p_x + i p_y) & -m\mp |\Delta|-Bp^2\end{array}\right),
\label{EQ:PHam}
\end{equation}
in the $\frac{1}{\sqrt{2}}(c_{{\bf p}\uparrow}+ c^\dagger_{-{\bf p}\downarrow}, c_{{\bf p}\downarrow}+c^\dagger_{-{\bf p}\uparrow}, -c_{{\bf p}\uparrow}+c^\dagger_{-{\bf p}\downarrow}, -c_{{\bf p}\downarrow}+c^\dagger_{-{\bf p}\uparrow})^T$ basis. The existence of Majorana edge states due to $h_\pm$ depends entirely on the sign of $m\pm|\Delta|$~\cite{READ2000, BERNEVIG2006, Qi2010} as Eq.~\eqref{EQ:PHam} is identical to the Hamiltonian of a $p_x + ip_y$ superconductor~\cite{READ2000}. Thus, $|\Delta|<-m$ gives two chiral Majorana edge states, $|\Delta|>|m|$ a single chiral Majorana edge state, and $|\Delta|<m$ no edge states. For general $\mu$, the condition reads~\cite{Qi2010}
\begin{equation}
\begin{array}{cc} m<-\sqrt{|\Delta|^2 + \mu^2}  & \Rightarrow \mathcal{N}=2,\\
m^2 < |\Delta|^2 + \mu^2 & \Rightarrow \mathcal{N}=1,\\
m>\sqrt{|\Delta|^2 + \mu^2} & \Rightarrow \mathcal{N}=0,\end{array}
\label{EQ:ChernCond}
\end{equation}
as the bulk gap closes only at boundaries between these three cases. These conditions show that opening up an infinitesimal SC gap gives us the $\mathcal{N}=2$ TSC if the normal state is in the QAH phase, but it gives us the $\mathcal{N}=1$ TSC if the normal state is a metal ($|\mu|>|m|$) obtained from doping the QAH system. This doping can come from the Fermi level mismatch between the QAH insulator and the SC used to induce the proximity effect. In the following we consider a QAH bar in proximity with a SC island in the middle, as shown in Fig.~\ref{FIG:QAH-TSC backscatter}. The superconducting region can have different topological invariant determined by the conditions in Eq.~\eqref{EQ:ChernCond}. Here, we will only consider the cases where the SC island is in a uniform topological phase.

The effect of the SC island on the edge state is determined by the topological invariant of the SC island. 
We first note that, in the basis of Eq.~\eqref{EQ:zeroChem} and \eqref{EQ:PHam}, the QAH edge state can be considered as two identical copies of chiral Majorana fermions, one from the upper block of Eq.~\eqref{EQ:zeroChem} being $e+h$ and the other from the lower block being $e-h$~\cite{Fu2008, Qi2010}, where $e$ and $h$ denote particle and hole states, respectively. In this sense, the $\mathcal{N}=2$ TSC is topologically equivalent to the QAH insulator, and at the boundary between these two phases there will be no chiral state, as there is no change in the signs of the gaps in both $h_\pm$ terms. Therefore, if we have the $\mathcal{N}=2$ TSC for the SC island (the top of Fig.~\ref{FIG:QAH-TSC backscatter}), the edge current will be perfectly transmitted. By contrast, the NSC is by definition topologically trivial and supports no edge states. This also implies that there will be a chiral edge state at the boundary between the QAH region and the NSC. Therefore, if we have NSC for the SC island (the middle of Fig.~\ref{FIG:QAH-TSC backscatter}), there will be complete backscattering. In both cases, the edge state will not be involved in any violation of particle number conservation despite presence of the SC island. In the case of the $\mathcal{N}=1$ TSC with $\mu=0$, we can see from Eq.~\eqref{EQ:zeroChem} and \eqref{EQ:PHam} that, as $|\Delta|>|m|$, the single chiral Majorana edge state comes from $h_-$. However, this also implies that at the boundary between a $\mathcal{N}=1$ TSC and the QAH region, $m+|\Delta|$ (the gap of $h_+$) should change sign, meaning that there is a chiral Majorana state at this boundary~\cite{READ2000, BERNEVIG2006}. Therefore, when the $\mathcal{N}=1$ TSC region is inserted in the middle of a QAH bar (the bottom of Fig.~\ref{FIG:QAH-TSC backscatter}), the chiral edge state of the QAH region fractionalizes into two well-separated chiral Majorana states, in a manner analogous to the topological insulator surface state in proximity with a ferromagnet and a SC~\cite{FU2009}. The interesting point in the setup of Fig.~\ref{FIG:QAH-TSC backscatter} is that while one branch of chiral Majorana fermions is perfectly transmitted, the other branch is perfectly reflected.

For this reason, in the $\mathcal{N}=1$ TSC setup of Fig.~\ref{FIG:QAH-TSC backscatter} there is equal probability for normal and Andreev scattering. To show this, we need to obtain the $S$-matrix $s_{ij;\alpha\beta}$ of the TSC region which relates the annihilation operators for incoming modes $\hat{a}_{j;\beta}$ to the annihilation operators for outgoing modes $\hat{b}_{i;\alpha}$,
\begin{equation}
\hat{b}_{i;\alpha} = s_{ij;\alpha\beta} \hat{a}_{j;\beta},
\label{EQ:SDef}
\end{equation}
where $i=1,2$ is the lead label and $\alpha,\beta = e,h$ is the particle/hole label, which means $\hat{a}_{i;h}= \hat{a}_{i;e}^\dagger, \hat{b}_{i;h} = \hat{b}_{i;e}^\dagger$. Since the QAH edge state splits into two chiral Majorana edge states, the $S$-matrix can be factorized into two parts,
\begin{equation}
\left(\begin{array}{c} \hat{b}_{1;e}+\hat{b}_{1;e}^\dagger \\  \hat{b}_{2;e}+\hat{b}_{2;e}^\dagger \end{array}\right)\!=\!\left(\begin{array}{cc} -1 & 0 \\  0 & 1 \end{array} \right)\!\left(\begin{array}{c} \hat{a}_{1;e}+\hat{a}_{1;e}^\dagger \\  \hat{a}_{2;e}+\hat{a}_{2;e}^\dagger \end{array}\right),
\label{EQ:scatterMajorana}
\end{equation}
and
\begin{equation}
\left(\begin{array}{c} \hat{b}_{1;e}-\hat{b}_{1;e}^\dagger \\  \hat{b}_{2;e}-\hat{b}_{2;e}^\dagger \end{array}\right) =  \left(\begin{array}{cc} 0 & 1\\ 1 & 0\end{array}\right)\left(\begin{array}{c} \hat{a}_{1;e}-\hat{a}_{1;e}^\dagger \\  \hat{a}_{2;e}-\hat{a}_{2;e}^\dagger \end{array}\right).
\label{EQ:transMajorana}
\end{equation}
This gives the $S$-matrix
\begin{equation}
\left(\begin{array}{c} \hat{b}_{1;e} \\ \hat{b}_{1;e}^\dagger \\ \hat{b}_{2;e} \\ \hat{b}_{2;e}^\dagger   \end{array}\right) = \frac{1}{2} \left(\begin{array}{cccc} -1 & -1 & 1 & - 1 \\ -1 & -1 & - 1 & 1 \\ 1 & - 1 & 1 & 1 \\ - 1& 1 & 1 & 1 \end{array} \right) \left(\begin{array}{c} \hat{a}_{1;e} \\ \hat{a}_{1;e}^\dagger \\ \hat{a}_{2;e} \\ \hat{a}_{2;e}^\dagger \end{array}\right).
\label{EQ:majoranaS}
\end{equation}
The elements of the first column of this $S$-matrix correspond to the amplitudes for normal reflection, Andreev reflection, normal transmission, and Andreev transmission for the incoming right-moving particle, respectively; they are all equal up to signs. Likewise, the third column shows that these four scattering processes have equal probability for the incoming left-moving particle. The crucial point is that only when backscattering is due to the $\mathcal{N}=1$ TSC does it involve Andreev scattering. Backscattering from the $\mathcal{N}=2$ TSC or NSC does not involve Andreev scattering, as the splitting of the QAH chiral edge state into two chiral Majorana edge states does not happen in these cases. 

\begin{figure}[t]
\centerline{\includegraphics[width=.39\textwidth]{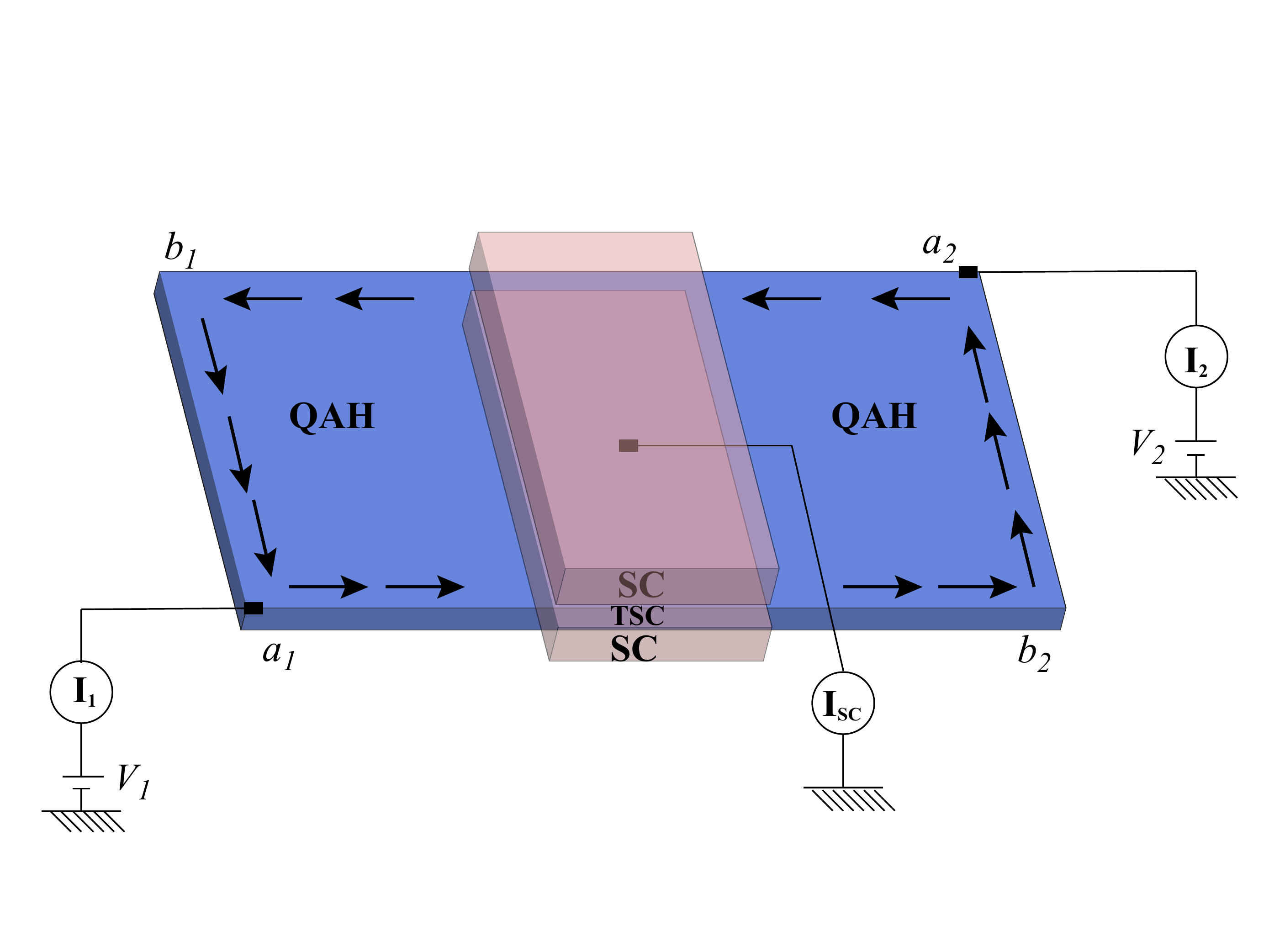}}
\caption{Schematic of the backscattering setup. The voltages 
$V_1$ and $V_2$ are applied to 
the lead 1 and 2, respectively, and set the chemical potential for the incoming modes that are annihilated by $\hat{a}_{1;e}$ and $\hat{a}_{2;e}$. In the central region, TSC is induced through proximity to two SC slabs that sandwich the QAH. The TSC is grounded through a contact to its bulk.}
\label{FIG:setup}
\end{figure}

Because of this property, changes of current due to the incoming modes $a_1, a_2$ of Fig.~\ref{FIG:setup} do not lead to changes of current due to the outgoing modes $b_1, b_2$ for the $\mathcal{N}=1$ TSC. To analyze this setup using the above $S$-matrix, we use Anantram and Datta's generalization~\cite{ANANTRAM1996} of the Landauer-B\"{u}ttiker formalism which allows for Andreev scattering. In this formalism, the conventional Landauer-B\"{u}ttiker formula~\cite{BUTTIKER1992} is extended by adding a particle/hole index,
\begin{eqnarray}
\hat{I}_i &=& \frac{e}{h}\int dE \sum_\alpha {\rm sgn}(\alpha) (\hat{a}^\dagger_{i;\alpha} \hat{a}_{i;\alpha}-\hat{b}^\dagger_{j;\beta} \hat{b}_{k;\gamma})\nonumber\\
&=& \frac{e}{h}\int dE \sum_\alpha \sum_{jk;\beta\gamma} {\rm sgn}(\alpha) A^{(i\alpha)}_{jk;\beta\gamma} \hat{a}^\dagger_{j;\beta} \hat{a}_{k;\gamma},
\label{EQ:Datta}
\end{eqnarray}
where 
${\rm sgn}(e) = 1$, ${\rm sgn}(h) = -1$, 
and, to make Eq.~\eqref{EQ:Datta} consistent with Eq.~\eqref{EQ:SDef}, $A^{(i\alpha)}$ is a $4 \times 4$ matrix defined as
\begin{equation}\nonumber
A^{(i\alpha)}_{jk;\beta\gamma} \equiv \delta_{ij} \delta_{ik} \delta_{\alpha\beta}\delta_{\alpha\gamma} - s^*_{ij;\alpha\beta}s_{ik;\alpha\gamma}.
\end{equation}
Just as in the Blonder-Tinkham-Klapwijk formula for the conductance of a normal-superconducting interface~\cite{BLONDER1982}, contributions from Andreev scattering cancel those of normal scattering. This gives us the incoming currents for 
the leads 1 and 2~\cite{ANANTRAM1996, ENTIN-WOHLMAN2008},
\begin{align}
I_1 &= \frac{e^2}{h}[(1\!-\!\mathcal{R}\!+\!\mathcal{R}_A)(V_1\!-\!V_{SC})-(\mathcal{T}\!-\!\mathcal{T}_A)(V_2\!-\!V_{SC})],\nonumber\\
I_2 &=\frac{e}{h}[(1\!-\!\mathcal{R}\!+\!\mathcal{R}_A)(V_2\!-\!V_{SC})-(\mathcal{T}\!-\!\mathcal{T}_A)(V_1\!-\!V_{SC})],
\label{EQ:current}
\end{align}
where $V_{SC}$ is the voltage applied to the SC island, $\mathcal{R}, \mathcal{T}$ are normal reflection and transmission probabilities and $\mathcal{R}_A, \mathcal{T}_A$ are Andreev reflection and transmission probabilities for the incoming particles.


According to the first and the third column of Eq.~\eqref{EQ:majoranaS}, we have $\mathcal{R}=\mathcal{R}_A=\mathcal{T}=\mathcal{T}_A=1/4$. Thus, we see that the setup of Fig.~\ref{FIG:setup} measures a half-integer Hall conductivity for the $\mathcal{N}=1$ case of Fig.~\ref{FIG:QAH-TSC backscatter}. When we ground the SC island ($V_{SC} =0$) and apply voltage $V_1 = -V_2 = V/2$, Eq.~\eqref{EQ:current} gives
\begin{equation}\nonumber
I_1 = -I_2 = \frac{e^2}{2h}V,
\end{equation}
which are currents flowing 
into the lead 1 and 2, respectively. 
This result is due to the increase of incoming current by $e^2 V/2h$ in the lead 1 and decrease of incoming current by $e^2 V/2h$ in the lead 2, while outgoing current stays the same for both leads. Thus, charge is conserved in this voltage setup. In addition, the QAH edge is grounded along with the SC region and all the voltage drop occurs at the contacts to the voltage. By contrast, if we consider the $\mathcal{N}=2$ TSC or the NSC in the same setup, 
we will measure Hall conductivity of $e^2/h$ and $0$, respectively.

Whether Andreev scattering processes are involved or not can be shown directly by measuring the current flowing from the SC island to the ground. When $V_2 \neq -V_1$, the change in the right-moving current $e^2 V_1 /h$ will not be canceled by the change in the left-moving current $e^2 V_2 /h$. In the case of the $\mathcal{N}=1$ TSC however, these currents do not flow out into the outgoing modes $b_1, b_2$ of the QAH region. Rather, since the SC island in our setup is not floating but grounded (Fig.~\ref{FIG:setup}), the net incoming current
\begin{equation}
I_{SC} = \frac{e^2}{h}(V_1 + V_2),
\label{EQ:SCground}
\end{equation}
will be flowing from the SC island to the ground. On the other hand, for the NSC or the $\mathcal{N}=2$ TSC, all incoming current will flow out into the outgoing modes $b_1, b_2$ of the QAH region, and no current will flow from the SC island to the ground. This will hold true regardless of whether we have $V_2 \neq -V_1$ or not. In other words, whereas the current is solely determined by $V_1 - V_2$ for the NSC or the $\mathcal{N}=2$ TSC, the same cannot be said for the $\mathcal{N}=1$ TSC.

Noise measurements will also show qualitative differences from the case of normal backscattering, because the correlation between current fluctuations in the same lead is unaffected by backscattering while the correlation between current fluctuations in different leads vanishes. From the Anantram-Datta formalism, we obtain the zero frequency noise~\cite{ANANTRAM1996},
\begin{eqnarray}
S_{ij} &=& \frac{2e^2}{h} \sum_{\alpha\beta} \sum_{kl; \gamma\delta} \int dE {\rm sgn}(\alpha){\rm sgn}(\beta) A^{(i\alpha)}_{kl;\gamma\delta} A^{(j\beta)}_{lk;\delta\gamma}\nonumber\\
&\times& n_{k\gamma}(1-n_{l\delta}).\nonumber
\end{eqnarray}
This formalism gives us the following general formula for the thermal noise, which applies for $eV_{1,2} \ll k_B T$ at temperature $T$:
\begin{eqnarray}
S_{11} = S_{22} &=& \frac{4e^2}{h}k_B T(1-\mathcal{R}+\mathcal{R}_A),\nonumber\\
S_{12} &=& \frac{4e^2}{h}k_B T(-\mathcal{T}+\mathcal{T}_A),
\label{EQ:noiseScatter}
\end{eqnarray}
which clearly gives $S_{11} = S_{22} = 4(e^2/h)k_B T$ and $S_{12}=0$. Interestingly, even if the $\mathcal{N}=1$ TSC is near the transition to the $\mathcal{N}=2$, so that the $e-h$ chiral Majorana state is not reflected perfectly as in Eq.~\eqref{EQ:scatterMajorana}, we would still have $\mathcal{R}=\mathcal{R}_A$ as long as Eq.~\eqref{EQ:transMajorana} stays unmodified. Thus in this case, although $S_{12} \neq -S_{ii}$, we do have $S_{11} = S_{22} = 4(e^2/h)k_B T$ just like the case where the chiral edge state transmits perfectly. Similarly, near the transition between $\mathcal{N}=1$ TSC and $\mathcal{N}=0$ NSC, we have $S_{12}=0$ just like the case chiral edge state reflects perfectly, though $S_{11}$ and $S_{22}$ are nonzero.

We emphasize that the above results for the thermal noise hold only because we are keeping the chemical potential of the SC fixed by grounding it. If the SC potential is floating, that is, not connected directly to any external voltage, $V_{SC}$ will reach a value where $I_1 + I_2 =0$, which would conserve the charge of the SC. However, this charge conservation should also be applied to current fluctuations, which means that $\mu_\mathrm{SC}$ should fluctuate with $\delta I_1$, $\delta I_2$. Such $\mu_\mathrm{SC}$ fluctuations will result in the Johnson-Nyquist relation $S_{11}=S_{22}=-S_{12} = 4G k_B T$ where $G$ is the conductance between the two leads, which should always hold for the two-terminal measurement~\cite{ANANTRAM1996}.

These subtleties do not have any effect when backscattering is caused by the $\mathcal{N}=2$ TSC or NSC, where $S_{11}=S_{22}=-S_{12} = 4G k_B T$ always holds. This is because in those cases $\mathcal{R}+\mathcal{T}=1$ and $\mathcal{T}_A = \mathcal{R}_A = 0$. For the setup of Fig.~\ref{FIG:QAH-TSC backscatter}, we have $\mathcal{R}=1$ for the NSC, which gives $G=0$ and thus reduces all noise to zero. For the $\mathcal{N}=2$ TSC, we have $\mathcal{R}=1$ for the NSC, which gives $G=e^2/h$ and $S_{11}=S_{22}=-S_{12} = 4(e^2/h) k_B T$. Whereas for the conductance, the $\mathcal{N}=1$ TSC gives a value that is halfway between that of the NSC and the $\mathcal{N}=2$ TSC, for the noise, we find $S_{ii}$ to be just that of the $\mathcal{N}=2$ TSC and $S_{12}$ to be that of the NSC.

In summary, we have described a method to detect a chiral Majorana state through transport measurements. Probabilities for normal and Andreev scattering are equal when backscattering occurs through a single chiral Majorana state. If the voltage is applied symmetrically ($V_2 = -V_1$),  this gives a half-integer Hall conductivity, and if not, we will have a quantized current flowing from the SC to the ground [Eq.~\eqref{EQ:SCground}]. Also, due to normal and Andreev scattering having the same probability, the correlation of current fluctuations within the same lead 
is unaffected by backscattering, while the correlation of current fluctuations in different leads vanishes.

We would like to thank Steve Kivelson, David Goldhaber-Gordon, Mac Beasley, Matthew Fisher and Liang Fu for sharing their insights. This work is supported by 
DOE under contract DE-AC02-76SF00515 (SBC), the Stanford Graduate Fellowship Program (JM) and 
NSF under grant number DMR-0904264.


\end{document}